\pdfoutput=1 
\documentclass
[preprint,pra,amsfonts,amssymb,10pt,showpacs,showkeys,byrevtex,onecolumn]{revtex4}%
\usepackage{amsfonts}
\usepackage{amsmath}
\usepackage{amssymb}
\usepackage{graphicx}%
\providecommand{\U}[1]{\protect\rule{.1in}{.1in}}

\begin{document}
\preprint{PHYSICS/123-qed}
\title[Relaxation of spins in gradient]{Relaxation of spins due to a magnetic field gradient, revisited; Identity of
the Redfield and Torrey theories}
\author{R. Golub}
\affiliation{Physics Department, North Carolina State University, Raleigh, NC 27695\ \ \ \ \ \ \ \ \ \ \ \ \ \ \ \ \ \ \ \ \ \ \ \ \ \ \ \ \ \ }
\author{Ryan M. Rohm}
\affiliation{\ Physics Department, University of North Carolina, Chapel Hill, NC 27599}
\author{C.M. Swank}
\affiliation{Physics Department, North Carolina State University, Raleigh, NC 27695}

\begin{abstract}
There is an extensive literature on magnetic gradient induced spin relaxation.
Cates, Schaefer and Happer (CSH, \cite{CSH}), in a seminal paper, have solved
the problem in the regime where diffusion theory (the Torrey equation
\cite{Tor}) is applicable using an expansion of the density matrix in
diffusion equation eigenfunctions and angular momentum tensors. McGregor
\cite{McG} has solved the problem in the same regime using a slightly more
general formulation using Redfield theory formulated in terms of the
auto-correlation function of the fluctuating field seen by the spins and
calculating the correlation functions using the diffusion theory Green's
function. The results of both calculations were shown to agree for a special
case, \cite{McG}. In the present work we show that the eigenfunction expansion
of the Torrey equation yields the expansion of the Green's function for the
diffusion equation thus showing the identity of this approach with that of
Redfield theory. The general solution can also be obtained directly from the
Torrey equation for the density matrix. Thus the physical content of the
Redfield and Torrey approaches are identical. We then introduce a more general
expression for the position autocorrelation function of particles moving in a
closed cell,  extending the range of applicability of the theory.

\end{abstract}
\volumeyear{year}
\volumenumber{number}
\issuenumber{number}
\eid{identifier}
\date[Date text]{date}
\received[Received text]{date}

\revised[Revised text]{date}

\accepted[Accepted text]{date}

\published[Published text]{date}

\startpage{101}
\endpage{102}
\maketitle
\tableofcontents

\section{Introduction}

The problem of relaxation in nuclear magnetic resonance due to field gradients
has been discussed by many authors but continues to be a topic of current
research. Recently attention has been focussed on this subject in connection
with searches for new P,T violating forces mediated by the hitherto unobserved
Axion \cite{Petu}, \cite{poko}.

We give a short, very incomplete, summary of how the field developed until
now. In 1950 Hahn \cite{Hah} used his just invented spin echo technique to
study the effect of translational diffusion on relaxation in nmr. Torrey, in
1953 \cite{Tor2} gave a derivation of the effect of translational diffusion
that had been alluded to by Hahn. In 1954 Carr and Purcell \cite{CarrP}
presented a more elaborate method for measuring diffusion constants using
relaxation due to translational diffusion in an inhomogeneous field with known
gradient. Then in 1956 Torrey \cite{Tor} introduced a specific partial
differential equation (Torrey equation) describing the effects of diffusion on
relaxation. He showed that under conditions when diffusion theory was valid
the physics was described by adding a diffusion term to the usual Bloch
equations. These treatments of diffusion did not take into account the effect
of the boundaries of the measurement cell. Ten years later (1966) Baldwin
Robertson \cite{ROB} gave an approximate solution of the Torrey equation in a
relatively small region, defined by 2 parallel planes, where the influence of
the boundaries was important. Using the method of phase accumulation and
assuming the phase distribution to be Gaussian, Neuman, in 1973 \cite{Neum}
gave an approximate solution for planar, cylindrical and spherical geometries
and showed this was in agreement with Robertson's results.

In 1987, Cates, Schaefer and Happer (CSH) \cite{CSH} calculated the relaxation
for parameters where the diffusion theory is appropriate using second order
perturbation theory and an expansion in eigenfunctions of the Torrey equation
applied to  the density matrix. At high densities the perturbation theory
breaks down and at low densities ($\lambda\gtrsim R)$ the diffusion theory is
invalid. (The present work shows how to go beyond this latter limit.) The
authors start with the equation of motion for the density matrix in the
presence of diffusion, (Torrey equation) \cite{Tor}%
\begin{equation}
\frac{\partial\rho}{\partial t}=\frac{1}{i\hbar}\left[  H,\rho\right]
+D\nabla^{2}\rho\label{A}%
\end{equation}
They consider the deviations of the magnetic field from the volume averaged
field as a perturbation and the volume averaged field (taken along $z$) as the
unperturbed system. Then, expanding $\rho$ in the 'eigenpolarizations' of the
unperturbed problem and carrying out a perturbation expansion in the field
variation, taken to be varying linearly with position, they obtain a solution
valid to second order in the perturbation.

McGregor \cite{McG} has given a slightly more general treatment  based on
Redfield's relaxation matrix theory \cite{Red} , as presented by Slichter
\cite{Sli}. The starting point of this treatment is the equation of motion for
the density matrix expanded to second order in a perturbation (\cite{Sli},
equation 5.313)
\begin{equation}
\frac{\partial\rho^{\ast}}{\partial t}=\frac{1}{i\hbar}\left[  H_{1}^{\ast
}\left(  t\right)  ,\rho^{\ast}\left(  0\right)  \right]  +\left(  \frac
{i}{\hbar}\right)  ^{2}\int_{0}^{t}\left[  \left[  \rho^{\ast}\left(
0\right)  ,H_{1}^{\ast}\left(  t^{\prime}\right)  \right]  ,H_{1}^{\ast
}\left(  t\right)  \right]  dt^{\prime}%
\end{equation}
where $H_{1}^{\ast}$ represents the deviation of the field from its volume
averaged value and starred quantities are expressed in the interaction
representation with the volume average field considered as the unperturbed system.

The results show that the relaxation depends on the auto-correlation function
of the fluctuating field (frequency spectrum of the field fluctautions) as
seen by the spins as they move through the measurement cell and the
correlation function is determined by the diffusion theory Green's function
for the case when diffusion theory is valid.

For high densities, when the boundary conditions do not play a role, the exact
solution obtained by Torrey \cite{Tor} is valid.

Following this work in 1991, Stoller, Happer and Dyson \cite{STHappDyson}%
\ have shown how to use the exact eigenfunctions of the Torrey equation (Airy
functions) to get exact solutions in one dimension. de Swiet and Sen
\cite{deSwSen} have used this and other approaches to study a wider range of
geometries. Hayden et al (2004) give a nice discussion of the Gaussian phase
distribution work along with experimental confirmation in a cylindrical
geometry \cite{MHayd}.

McGregor \cite{McG} has shown that the results of his Redfield theory
treatment are equivalent to those obtained from the Torrey equation \cite{CSH}
for the special case of the high pressure limit in a spherical cell.
Nevertheless it is illuminating to note that the expansion in the diffusion
equation eigenfunctions obtained by CSH \cite{CSH} is in fact the usual
eigenfunction expansion of the Green's function and hence the results based on
the Torrey equation \cite{CSH} and those of the Redfield theory \cite{McG} are
identical for all cases considered by CSH. We show this in the next section,
with details confined to an appendix. Thus the physical content of the two
approaches are identical in spite of their rather different starting points.

We then show how these results can be applied beyond the diffusion theory
limits by giving an analytic expression for the trajectory correlation
functions valid for a range of pressures wider than that for which diffusion
theory is applicable.

\section{Equivalence of the Torrey equation and Redfield theory results when
diffusion theory is valid.}

In the appendix we review the calculation of CSH applied to spin 1/2 and using
a slightly altered notation. We expand the density matrix in the spin 1/2
operators, $\sigma_{0,\pm1}.$

The result for $T_{1}$, equation (\ref{111a}), compare equation (50), CSH:%
\begin{equation}
\frac{1}{T_{1}}=\frac{4}{V}\operatorname{Re}\int\int\left[  \Omega_{1}\left(
\overrightarrow{r^{\prime}}\right)  \right]  _{-}\left[  \Omega_{1}\left(
\overrightarrow{r}\right)  \right]  _{+}\sum_{\beta^{\prime}}\ \left(
\frac{\phi_{\beta^{\prime}}\left(  \overrightarrow{r}^{\prime}\right)
\phi_{\beta^{\prime}}\left(  \overrightarrow{r}\right)  }{(Dk_{\beta^{\prime}%
}^{2}-i2\Omega_{o})}\right)  d^{3}r^{\prime}d^{3}r
\end{equation}
is seen to contain the Fourier transform of the eigenfunction expansion of the
Green' function, equation (\ref{1a} )%
\begin{equation}
\widetilde{G}\left(  \overrightarrow{r},\overrightarrow{r}^{\prime}%
,\omega\right)  =\sum_{\beta^{\prime}}\ \left(  \frac{\phi_{\beta^{\prime}%
}\left(  \overrightarrow{r}^{\prime}\right)  \phi_{\beta^{\prime}}\left(
\overrightarrow{r}\right)  }{(Dk_{\beta^{\prime}}^{2}-i\omega)}\right)
\label{01}%
\end{equation}
so that we have (equation \ref{T1}, eqn.9 in \cite{McG})%
\begin{equation}
\frac{1}{T_{1}}=\frac{\gamma^{2}}{2}\int_{-\infty}^{\infty}d\tau
e^{i\omega_{o}\tau}\left\langle \left[  B_{1}\left(  t\right)  \right]
_{x}\left[  B_{1}\left(  t+\tau\right)  \right]  _{x}+\left[  B_{1}\left(
t\right)  \right]  _{y}\left[  B_{1}\left(  t+\tau\right)  \right]
_{y}\right\rangle
\end{equation}

Similarly the results for $T_{2}$ (\ref{XX}) are also equivalent to McGregor's
results (eqn. 10 in \cite{McG}) when we take (\ref{01}) in the form%

\begin{equation}
\sum_{\beta^{\prime}}\ \left(  \frac{\phi_{\beta^{\prime}}\left(
\overrightarrow{r}^{\prime}\right)  \phi_{\beta^{\prime}}\left(
\overrightarrow{r}\right)  }{Dk_{\beta^{\prime}}^{2}}\right)  =\widetilde
{G}\left(  \overrightarrow{r},\overrightarrow{r}^{\prime},\omega=0\right)
=\int_{0}^{\infty}d\tau G\left(  \overrightarrow{r},t|\overrightarrow
{r}^{\prime},t^{\prime}\right)
\end{equation}

\subsection{Direct solution using Green's function}

As we have shown that the CSH result in terms of diffusion equation
eigenfunctions is identical with the McGregor result using the Redfield theory
and the diffusion theory Green's function it should be possible to derive the
result starting with the Torrey equation, (\ref{A}) (equation (\ref{2}) in
Appendix A) using (\ref{01a})%
\begin{equation}
\frac{\partial\rho}{\partial t}=\frac{1}{i}\Gamma_{0}\rho+\frac{\eta}{i}%
\Gamma_{1}\rho+D\triangledown^{2}\rho
\end{equation}
We expand $\rho$ as in (\ref{5})%
\begin{equation}
\rho\left(  \overrightarrow{r},t\right)  =\sum_{j}\sigma_{j}f_{j}\left(
\overrightarrow{r},t\right)
\end{equation}
taking the trace with $\sigma_{i}^{T}$obtaining
\begin{equation}
\frac{\partial f_{i}^{\prime}}{\partial t}-D\nabla^{2}f_{i}^{\prime}=-\frac
{i}{\alpha_{i}}\sum_{j}\left[  \Gamma_{1}\right]  _{ij}f_{j}^{\prime
}e^{i\left(  \Lambda_{i}-\Lambda_{j}\right)  t}\label{0011}%
\end{equation}
with%
\begin{align}
f_{i} &  =f_{i}^{\prime}(x,t)e^{-i\Lambda_{i}t}\\
\Lambda_{i} &  =2\Omega_{o}M_{i}%
\end{align}

We will treat the sum on the r.h.s. as a perturbation introducing the Green's
function for the unperturbed problem, $G_{0}\left(  x,\tau\right)  ,$
satisfying
\begin{equation}
\frac{\partial G_{0}\left(  x,t\right)  }{\partial t}-D\nabla^{2}G_{0}\left(
x,t\right)  =\delta^{\left(  3\right)  }\left(  x\right)  \delta\left(
t\right)  \label{0011a}%
\end{equation}
and the boundary condition
\begin{equation}
\overrightarrow{n}\cdot\overrightarrow{\triangledown}G_{0}=0
\end{equation}

Then we can convert (\ref{0011}) to an integral equation for $f_{i}^{\prime}$
\begin{equation}
f_{i}^{\prime}(x,t)=f_{i}^{\prime(0)}+\int G_{0}\left(  x-x^{\prime
},t-t^{\prime}\right)  \frac{1}{i\alpha_{i}}\sum_{j}\left[  \Gamma_{1}\left(
x^{\prime}\right)  \right]  _{ij}f_{j}^{\prime}\left(  x^{\prime},t^{\prime
}\right)  e^{i\left(  \Lambda_{i}-\Lambda_{j}\right)  t^{\prime}}dx^{\prime
}dt^{\prime}%
\end{equation}
which can be solved by iteration ($f_{i}^{\prime(0)}$ being a solution of
(\ref{0011}) with the r.h.s. set equal to 0)%
\begin{align}
f_{i}^{\prime}(x,t) &  =f_{i}^{\prime(0)}+\int G_{0}\left(  x-x^{\prime
},t-t^{\prime}\right)  \frac{1}{i\alpha_{i}}\sum_{j}\left[  \Gamma_{1}\left(
x^{\prime}\right)  \right]  _{ij}f_{j}^{\prime(0)}\left(  x^{\prime}%
,t^{\prime}\right)  e^{i\left(  \Lambda_{i}-\Lambda_{j}\right)  t^{\prime}%
}dx^{\prime}dt^{\prime}+...\nonumber\\
&  ..\int\int G_{0}\left(  x-x^{\prime},t-t^{\prime}\right)  \frac{1}%
{i\alpha_{i}}\sum_{j}\left[  \Gamma_{1}\left(  x^{\prime}\right)  \right]
_{ij}e^{i\left(  \Lambda_{i}-\Lambda_{j}\right)  t^{\prime}}G_{0}\left(
x^{\prime}-x^{\prime\prime},t^{\prime}-t^{\prime\prime}\right)  \times
...\nonumber\\
&  ...\frac{1}{i\alpha_{j}}\sum_{k}\left[  \Gamma_{1}\left(  x^{\prime\prime
}\right)  \right]  _{jk}f_{k}^{\prime(0)}\left(  x^{\prime\prime}%
,t^{\prime\prime}\right)  e^{i\left(  \Lambda_{j}-\Lambda_{k}\right)
t^{\prime\prime}}dx^{\prime\prime}dt^{\prime\prime}dx^{\prime}dt^{\prime}%
\end{align}
If we now operate on this with $\partial/\partial t$ and use (\ref{0011a}),
noting that we will eventually integrate the result over $d^{3}x$ so that
terms containing $D\nabla^{2}G_{0}$ will vanish because of the boundary
condition, we find for the second order term:%
\begin{align}
\dot{f}_{i}^{\prime}(x,t) &  =\frac{1}{i\alpha_{i}}\sum_{j,k}\left[
\Gamma_{1}\left(  x\right)  \right]  _{ij}e^{i\left(  \Lambda_{i}-\Lambda
_{j}\right)  t}\int G_{0}\left(  x-x^{\prime\prime},t-t^{\prime\prime}\right)
\times...\nonumber\\
&  \frac{1}{i\alpha_{j}}\left[  \Gamma_{1}\left(  x^{\prime\prime}\right)
\right]  _{jk}e^{i\left(  \Lambda_{j}-\Lambda_{k}\right)  t^{\prime\prime}%
}f_{k}^{\prime(0)}\left(  x^{\prime\prime},t^{\prime\prime}\right)
dx^{\prime\prime}dt^{\prime\prime}%
\end{align}
and averaging over $d^{3}x,$ $\left\langle ...\right\rangle =\frac{1}{V}\int
d^{3}x\left(  ...\right)  $:%
\begin{align}
\left\langle \dot{f}_{i}^{\prime}(x,t)\right\rangle  &  =\frac{1}{Vi\alpha
_{i}}\int\int d^{3}xd^{3}x^{\prime\prime}dt^{\prime\prime}\sum_{j,k}\left[
\Gamma_{1}\left(  x\right)  \right]  _{ij}e^{i\left(  \Lambda_{i}-\Lambda
_{j}\right)  t}G_{0}\left(  x-x^{\prime\prime},t-t^{\prime\prime}\right)
\times...\nonumber\\
&  ...\frac{1}{i\alpha_{j}}\left[  \Gamma_{1}\left(  x^{\prime\prime}\right)
\right]  _{jk}e^{i\left(  \Lambda_{j}-\Lambda_{k}\right)  t^{\prime\prime}%
}f_{k}^{\prime(0)}\left(  x^{\prime\prime},t^{\prime\prime}\right)
\end{align}
To investigate relaxation we set $i=k.$  As we are interested in relaxation of
a spatially homogeneous gas we put $f_{i}^{\prime(0)}\left(  x^{\prime\prime
},t\right)  =f_{i}^{\prime(0)}\left(  t\right)  =const$ and take it out of the
integral since it is the solution of (\ref{0011}) with the r.h.s.=0. Then the
relaxation rate will be given by%
\begin{align}
\frac{\left\langle \dot{f}_{i}^{\prime}(x,t)\right\rangle }{f_{i}^{\prime
(0)}\left(  t\right)  } &  =-\frac{1}{V\alpha_{i}}\int\int d^{3}%
xd^{3}x^{\prime\prime}dt^{\prime^{\prime}}\sum_{j,k}\left[  \Gamma_{1}\left(
x\right)  \right]  _{ij}e^{i\left(  \Lambda_{i}-\Lambda_{j}\right)  t}%
G_{0}\left(  x-x^{\prime^{\prime}},t-t^{^{\prime}\prime}\right)
\times...\nonumber\\
&  ..\frac{1}{\alpha_{j}}\left[  \Gamma_{1}\left(  x^{\prime^{\prime}}\right)
\right]  _{jk}e^{i\left(  \Lambda_{j}-\Lambda_{k}\right)  t^{\prime\prime}}%
\end{align}
where $i=0$ will give $1/T_{1}$ and $i=1\left(  +\right)  $ will give
$1/T_{2}.$

Using equations (\ref{001a}, \ref{001b}, \ref{001c}, \ref{001d}) it is easy to
see that we obtain eqn.(\ref{111a}, \ref{111b}) for $1/T_{1}$ and (\ref{11},
\ref{111c}) for $1/T_{2}.$

Thus direct solution of the Torrey equation (\ref{2}) containing a diffusion
term, using the conventional second order perturbation theory based on the
Green's function for the unperturbed equation yields results in agreement with
those obtained by McGregor \cite{McG} by applying second order perturbation
theory to the equation of motion for the density matrix (Redfield theory),
where diffusion theory only enters through the  correlation functions of the
magnetic field and the physical content of the two theories is identical.

\section{Beyond Diffusion theory}

Having shown the equivalence of the CSH treatment based on the Torrey equation
to the calculation based on Redfield theory when diffusion theory is used in
evaluating the correlation functions, we widen the range of applicability by
introducing a form of the correlation function which is also valid when the
diffusion theory breaks down, i.e. when the condition $\lambda_{c}\ll L$ no
longer holds ($\lambda_{c}=v\tau_{c}$ is the collision mean free path and $L$
is a typical size of the containing vessel).

\subsection{Correlation functions for motion in a closed cell.}

Defining a correlation function as%
\begin{equation}
R_{fg}\left(  \tau\right)  =\left\langle f\left(  t\right)  g\left(
t+\tau\right)  \right\rangle
\end{equation}
with $\left\langle ...\right\rangle $ representing an ensemble and time
average, we have the following relations \cite{Pap}%
\begin{align}
R_{xv}\left(  \tau\right)   &  =\frac{d}{d\tau}R_{xx}\left(  \tau\right)
\nonumber\\
R_{vv}\left(  \tau\right)   &  =-\frac{d^{2}}{d\tau^{2}}R_{xx}\left(
\tau\right)  \label{011}%
\end{align}
so the determination of any one will determine the whole family.

Barabanov \emph{et al} \cite{Bar}\emph{ }have calculated the velocity
auto-correlation function for particles moving in a closed vessel with
specularly reflecting walls. The effect of gas collisions are taken into
account. The method was initially \cite{Bar} applied to cylindrical vessels
for a case where only the motion normal to the axis is relevant, and then to
rectangular shaped vessels \cite{GS}. The results have been checked by
numerical simulations for many cases \cite{Bar}, \cite{GS}, \cite{LamGo}. The
function $R_{xv}\left(  \tau\right)  $ obtained from $R_{vv}\left(
\tau\right)  $ by means of equation (\ref{011}), has been applied to the study
of a false electric dipole moment effect that arises in magnetic resonance
experiments in the presence of an electric field \cite{Bar}, \cite{LamGo}.
\ The result can easily be applied to spherical cavities, the only
modification being that the distribution of the angle, $\alpha$, (the angle
between the trajectory and the normal to the reflecting surface) will be
different in the case of a sphere. The correlation function, initially
obtained for a single velocity, can be averaged over the appropriate velocity distribution.

For simplicity we will concentrate on a rectangular vessel in this work. In
that case the motions in each of the 3 directions are independent \cite{GS},
so we concentrate on one dimension to begin. Equations (27, 36 and 37) of
\cite{Bar} can be combined to give (note the r.h.s. of (33) in that paper
should be set equal to unity),
\begin{equation}
R_{vv}\left(  \tau\right)  =\frac{8v_{i}^{2}}{\tau_{w}^{2}}%
{\displaystyle\sum\limits_{n=1,3,5..}}
\left[  \frac{\psi_{n}\left(  \tau\right)  }{\omega_{n}^{2}}\right]
\label{011b}%
\end{equation}
where \ the wall collision time, $\tau_{w}=2R\sin\alpha/v,$ for particles with
velocity $v$ (in the plane of the trajectory), moving in a cylinder or sphere
of radius $R$. For the rectangular case we take $\alpha=\pi/2$ and
$\ R=L_{i}/2,$ (the length of the cell along direction $x_{i}$) and then
$\tau_{w}=L_{i}/v_{i}$ for particles in a rectangular vessel, moving along
direction $x_{i}$ with velocity $v_{i}.$
\begin{align}
\omega_{n} &  =\frac{n\pi v_{i}}{L_{i}}\nonumber\\
\psi_{n}\left(  \tau\right)   &  =\frac{\eta_{1}e^{-\eta_{1}\tau}-\eta
_{2}e^{-\eta_{2}\tau}}{\eta_{1}-\eta_{2}}\label{011a}%
\end{align}
and
\begin{align}
\eta_{1,2} &  =\frac{1}{2\tau_{c}}\left(  1\pm s_{n}\right)  \\
s_{n} &  =\sqrt{1-4\omega_{n}^{2}\tau_{c}^{2}}%
\end{align}
with $\tau_{c,}$ the mean time between collisions. We see that $\omega_{n}%
\tau_{w}=n\pi$ so that
\begin{equation}
R_{vv}\left(  \tau\right)  =8v_{i}^{2}%
{\displaystyle\sum\limits_{n=1,3,5..}}
\left[  \frac{\psi_{n}\left(  \tau\right)  }{\left(  n\pi\right)  ^{2}%
}\right]
\end{equation}
and $R_{vv}\left(  0\right)  =v_{i}^{2}$ (\ref{pow2}).

Using equations (\ref{011}) we find
\begin{equation}
R_{xx}(\tau)=\frac{8}{\pi^{2}}v^{2}\tau_{c}%
{\displaystyle\sum\limits_{n=1,3,5..}}
\frac{1}{n^{2}s_{n}}\left[  \frac{e^{-\eta_{2}\tau}}{\eta_{2}}-\frac
{e^{-\eta_{1}\tau}}{\eta_{1}}\right]  \label{0111}%
\end{equation}
where the constant of integration has been chosen to satisfy $R_{xx}%
(\infty)=0$ and we see that (\ref{pow4})
\begin{equation}
R_{xx}(0)=\frac{8L^{2}}{\pi^{4}}%
{\displaystyle\sum\limits_{n=1,3,5..}}
\frac{1}{n^{4}}=\frac{L^{2}}{12}=\left\langle x^{2}\right\rangle
\end{equation}
in agreement with McGregor's result (\cite{McG}, eqn. 24) from diffusion theory.

If we introduce dimensionless time $\tau^{\prime}=\tau/\tau_{c}$ and note
that
\begin{align}
\omega_{n}\tau_{c} &  =\frac{n\pi}{l^{\prime}}\\
s_{n} &  =\sqrt{1-\left(  \frac{2n\pi}{l^{\prime}}\right)  ^{2}}%
\end{align}
with $l^{\prime}=L_{i}/\lambda_{c}$ where the collision mean free path,
$\lambda_{c}=v_{i}\tau_{c}$ we can write (\ref{0111}) as%
\begin{equation}
R_{xx}(\tau^{\prime})=\left(  \frac{L^{2}}{12}\right)  \frac{12\cdot16}%
{\pi^{2}l^{\prime2}}%
{\displaystyle\sum\limits_{n=1,3,5..}}
\frac{1}{n^{2}s_{n}}\left[  \frac{e^{-\left(  1-s_{n}\right)  \tau^{\prime}%
/2}}{\left(  1-s_{n}\right)  }-\frac{e^{-\left(  1+s_{n}\right)  \tau^{\prime
}/2}}{\left(  1+s_{n}\right)  }\right]
\end{equation}
Note that $s_{n}$ can be real or complex representing the transition between
diffusive and ballistic behavior. 

\bigskip Figure 1) shows a plot of $R_{xx}(\tau^{\prime})/R_{xx}(0)$ for
various values of $l^{\prime}$.

\subsection{Spectrum of the correlation functions.}

We start with the velocity auto-correlation function equation (\ref{011b}) and
take the Fourier transform of (\ref{011a}) using the definition of Fourier
integral used by McGregor \cite{McG}:%
\begin{equation}
\psi_{n}\left(  \omega\right)  =\int_{-\infty}^{\infty}\psi_{n}\left(
\tau\right)  e^{=i\omega\tau}d\tau
\end{equation}
so that%
\begin{equation}
\psi_{n}\left(  \omega\right)  =2\frac{\omega^{2}}{\tau_{c}}\frac{1}{\left(
\omega^{2}-\omega_{n}^{2}\right)  ^{2}+\omega^{2}/\tau_{c}^{2}}%
\end{equation}
and following (\ref{011b})%
\begin{equation}
\psi\left(  \omega\right)  =\frac{8v_{i}^{2}}{\tau_{w}^{2}}%
{\displaystyle\sum\limits_{n=1,3,5..}}
\left[  \frac{\psi_{n}\left(  \omega\right)  }{\omega_{n}^{2}}\right]
\end{equation}
Then the spectrum of the position auto-correlation function, $G_{xx}\left(
\omega\right)  $, which determines the relaxation is given by, following
(\ref{011})%

\begin{align}
G_{xx}\left(  \omega\right)   &  =\frac{\psi\left(  \omega\right)  }%
{\omega^{2}}=\frac{16v_{i}^{2}}{\tau_{w}^{2}\tau_{c}}%
{\displaystyle\sum\limits_{n=1,3,5..}}
\left[  \frac{1}{\omega_{n}^{2}\left[  \left(  \omega^{2}-\omega_{n}%
^{2}\right)  ^{2}+\omega^{2}/\tau_{c}^{2}\right]  }\right] \\
&  =\frac{16L^{4}}{\lambda v_{i}\pi^{6}}%
{\displaystyle\sum\limits_{n=1,3,5..}}
\frac{1}{n^{6}}\frac{1}{\left[  \left(  \frac{\omega^{\prime}}{n\pi}\right)
^{2}-1\right]  ^{2}+\left[  \frac{\omega^{\prime}l^{\prime}}{\left(
n\pi\right)  ^{2}}\right]  ^{2}} \label{0112a}%
\end{align}
where the last equation is written in terms of a normalized frequency,
$\omega^{\prime}=\omega L/v_{i}$ and length, $l^{\prime}=L/\lambda.$

Figure 2) shows $S\left(  \omega^{\prime},l^{\prime}\right)  =G_{xx}\left(
\omega\right)  /G_{xx}\left(  0\right)  $ as a function of $\omega^{\prime}$
with $l^{\prime}$ as a parameter. Note this is for a single velocity.
Averaging over the velocity distribution is straightforward.

Taking the limit of (\ref{0112a}) for $\omega^{\prime}<<1/\pi$ and
reintroducing $\omega,$ we find
\begin{equation}
G_{xx}\left(  \omega\right)  =16D_{1}%
{\displaystyle\sum\limits_{n=1,3,5..}}
\frac{1}{n^{6}}\frac{1}{\omega^{2}+\left(  \frac{\left(  n\pi\right)
^{2}D_{1}}{L^{2}}\right)  ^{2}}\label{001}%
\end{equation}
where $D_{1}=v_{i}\lambda$ is the diffusion constant for one dimension. This
is the Fourier transform of the diffusion theory Green's function for this
problem as obtained by McGregor \cite{McG}, eqn. (24). For high frequencies,
assuming $l^{\prime}$ large, we obtain (neglecting the $1$ in the denominator
in (\ref{0112a}):%
\begin{equation}
G_{xx}\left(  \omega\right)  =\frac{2v_{i}^{2}\tau_{c}}{\omega^{2}\left(
1+\omega^{2}\tau_{c}^{2}\right)  }\label{012}%
\end{equation}
which is identical to McGregor's eqn. (13) \cite{McG} for the high frequency
limit. In obtaining equation (\ref{012}) we assumed
\[
\frac{\omega L}{vn\pi}>>1
\]
which of course cannot hold for all $n$. This means we are not properly
accounting for the high $n$ modes, which in reality would have a contribution
of the form (\ref{001}), which is anyway small for large n, and is responsible
for the fact that (\ref{012}) is independent of the size of the vessel. See
the discussion under Fig. 3) in \cite{CSH}.

\section{Discussion}

The approaches of the two calculations are quite different. We have seen that
Cates, Schaefer and \ Happer \cite{CSH} solved \ the Torrey equation (\ref{1})
by assuming an exponential form for the time dependence of $\rho$ and
expanding the decay constant and amplitude in a power series in the
fluctuating field, treated as a perturbation. McGregor's approach is based on
the Redfield treatment of the equation of motion for the density matrix, eqn.
(\ref{1}) without the explicit introduction of a diffusion term. Recursion is
used to get a second order approximation to this equation and the second order
term is written in terms of the correlation functions of the fluctuating field
components as seen by the nuclei \cite{Sli}. The diffusion theory is then
introduced in the calculation of these correlation functions. Lastly we have
shown that the same results follow from the recursive expansion of the
integral equation, derived by use of the Green's function, in the manner of
the Born expansion.

Working out the details of the diffusion theory for a spherical cell McGregor
showed that his result is equivalent to that of \cite{CSH} in the high
pressure limit with Neuman boundary conditions. We have shown that the two
approaches give identical results whenever eqn. (\ref{1}) and the perturbation
theory is valid, thus clearing up any possible confusion as to when one or the
other of the two quite different approaches is valid. The physical content of
both theories is identical.

We have also presented a more general form of the position auto-correlation
function for the case of a rectangular cell which is valid beyond the region
of validity of diffusion theory.

\section{Acknowledgements}

We are grateful to Bradley Fillipone for a helpful remark.

\bigskip

\section{Appendix A}

\subsection{Perturbation theory of Cates, Schaefer and Happer \cite{CSH}.}

The authors start with the equation of motion for the density operator, $\rho
$, in diffusion approximation:%
\begin{equation}
\frac{\partial\rho}{\partial t}=\frac{1}{i\hslash}\left[  H,\rho\right]
+D\triangledown^{2}\rho\label{1}%
\end{equation}
The Hamiltonian $H$ is broken up into an main term, $H^{(0)}$ and a
perturbation $H^{(1)}:$%
\begin{align}
H  &  =H^{(0)}+H^{(1)}\\
H^{(0)}  &  =\hslash\Omega_{o}\sigma_{z}\\
H^{(1)}  &  =\eta\hslash\overrightarrow{\Omega}_{1}\cdot\overrightarrow
{\sigma}%
\end{align}
where $\Omega_{o}$ is chosen so that the volume average of $\overrightarrow
{\Omega}_{1}$ is zero and $\eta$ is an expansion parameter. (Note: normally
$\Omega_{o}=\gamma B_{o}$ so here $\Omega_{o,1\text{ }}$are 1/2 the usual
values, $\Omega_{o,1\text{ }}=\gamma B_{o,1}/2$)

We rewrite \ref{1} as%
\begin{equation}
\frac{\partial\rho}{\partial t}=\frac{1}{i}\left[  \Omega_{o}\sigma_{z}%
,\rho\right]  +\frac{\eta}{i}\left[  \overrightarrow{\Omega}_{1}%
\cdot\overrightarrow{\sigma},\rho\right]  +D\triangledown^{2}\rho\label{2}%
\end{equation}
We will approach the problem using time independent perturbation theory, that
is we substitute $\rho=\rho^{\prime}e^{-\gamma t}$ and obtain $\left(
\rho^{\prime}\neq f(t)\right)  $
\begin{equation}
0=\left(  \gamma+\frac{1}{i}\Gamma_{o}+D\triangledown^{2}+\frac{\eta}{i}%
\Gamma_{1}\right)  \rho^{\prime} \label{15}%
\end{equation}
where $\Gamma_{o,1}$ are linear operators (we now drop the prime on
$\rho^{\prime}$, using $\rho$ to indicate the time independent solution)%
\begin{align}
\Gamma_{o}\rho &  =\left[  \Omega_{o}\sigma_{z},\rho\right] \nonumber\\
\Gamma_{1}\rho &  =\left[  \overrightarrow{\Omega}_{1}\cdot\overrightarrow
{\sigma},\rho\right]  \label{01a}%
\end{align}

We will then expand $\rho$ in the spherical components of the spin 1/2
operators $\overrightarrow{\sigma}:$%
\begin{align}
\sigma_{\pm}  &  =\left(  \sigma_{x}\pm i\sigma_{y}\right)  /2=\sigma_{1,2}\\
\sigma_{0}  &  =\sigma_{z}%
\end{align}

\bigskip The $\sigma_{i}$ are seen to have the following properties%
\begin{align}
Tr(\sigma_{i}^{T}\sigma_{j})  &  =\delta_{ij}+\delta_{jo}\delta_{io}%
=\delta_{ij}\alpha_{j}\label{4}\\
\left[  \sigma_{0},\sigma_{i}\right]   &  =\left[  \sigma_{z},\sigma
_{i}\right]  =2M_{i}\sigma_{i}%
\end{align}
with $M_{1,2}=\pm1,M_{0}=0$, and
\begin{equation}
\alpha_{j}=\left\{
\begin{array}
[c]{c}%
1\qquad(j=1,2)\\
2\qquad(j=0)
\end{array}
\right\}
\end{equation}
Thus%

\begin{equation}
\Gamma_{o}\sigma_{i}=2\Omega_{o}M_{i}\sigma_{i} \label{a}%
\end{equation}

\bigskip We now follow CSH, \cite{CSH}, by introducing a perturbation
expansion for $\rho(\overrightarrow{r})$ and $\gamma$ into (\ref{15})$:$%
\begin{align}
\rho(\overrightarrow{r})  &  =\rho^{(0)}+\eta\rho^{(1)}+\eta^{2}\rho^{(2)}\\
\gamma &  =\gamma^{(0)}+\eta\gamma^{(1)}+\eta^{2}\gamma^{(2)}%
\end{align}
As this must hold for any value of $\eta$ we collect terms in equal powers of
$\eta:$%
\begin{align}
0  &  =\left(  \gamma^{(0)}+\frac{1}{i}\Gamma_{o}+D\triangledown^{2}\right)
\rho^{(0)}\label{16}\\
0  &  =\left(  \gamma^{(0)}+\frac{1}{i}\Gamma_{o}+D\triangledown^{2}\right)
\rho^{(1)}+\left(  \frac{1}{i}\Gamma_{1}+\gamma^{(1)}\right)  \rho
^{(0)}\label{17}\\
0  &  =\left(  \gamma^{(0)}+\frac{1}{i}\Gamma_{o}+D\triangledown^{2}\right)
\rho^{(2)}+\left(  \frac{1}{i}\Gamma_{1}+\gamma^{(1)}\right)  \rho
^{(1)}+\gamma^{(2)}\rho^{(0)} \label{18}%
\end{align}

\bigskip We look for a solution in the form
\begin{equation}
\rho^{(0)}=\sigma_{j}f_{j}^{(0)}(\overrightarrow{r}) \label{5}%
\end{equation}
Substituting into (\ref{16}) and applying $Tr(\sigma_{i}^{T}\cdot)$ to the
resultant equation yields%
\begin{equation}
\left(  \gamma_{i}^{(0)}-i2\Omega_{o}M_{i}+D\triangledown^{2}\right)
f_{i}^{(0)}(\overrightarrow{r})=0 \label{8a}%
\end{equation}
$f_{i}(\overrightarrow{r})$ has to satisfy boundary conditions on the surface
of the measurement cell. CSH, \cite{CSH},\ have taken the von Neuman
conditions (zero current at the walls) but as they point out the method can be
applied to the case where depolarization takes place at the walls. In any case
equation (\ref{8a}), along with the boundary conditions form an eigenvalue
problem. The solutions are given by the solution to
\begin{equation}
\left(  \triangledown^{2}+k_{\alpha}^{2}\right)  \phi_{\alpha}=0 \label{12}%
\end{equation}
where the eigenvalues $k_{\alpha}$ are determined by the boundary conditions.
Then (\ref{8a}) implies%
\begin{equation}
\gamma_{i,\alpha}^{(0)}=i2M_{i}\Omega_{o}+Dk_{\alpha}^{2} \label{13}%
\end{equation}

In order to solve for the higher order correction terms to the solution it is
useful to expand the corrections to $f_{i}(\overrightarrow{r})$ in a series of
the zero order functions, $f_{i\alpha}^{(0)}(\overrightarrow{r})=\phi
_{i\alpha},$ (the eigenfunctons of (\ref{12})).
\begin{equation}
\rho_{i\alpha}^{(n)}(\overrightarrow{r})=\sum_{j\beta^{\prime}}\sigma_{j}%
\phi_{\beta^{\prime}}a_{j\beta^{\prime},i\alpha}^{(n)}%
\end{equation}
which form a complete set of functions satisfying the boundary conditions.
$n=1$ or $2$ indicates the order of the correction. Thus (\ref{17}) becomes%
\begin{align*}
0  &  =\left(  \gamma_{i,\alpha}^{(0)}+\frac{1}{i}\Gamma_{o}+D\triangledown
^{2}\right)  \sum_{j,\beta^{\prime}}\sigma_{j}\phi_{\beta^{\prime}}%
a_{j\beta^{\prime},i\alpha}^{(1)}+\left(  \frac{1}{i}\Gamma_{1}+\gamma
_{i,\alpha}^{(1)}\right)  \sigma_{i}\phi_{\alpha}\\
0  &  =\sum_{j}\left(  \gamma_{i,\alpha}^{(0)}+\frac{1}{i}2M_{j}\Omega
_{o}+D\triangledown^{2}\right)  \sigma_{j}\sum_{\beta^{\prime}}\phi
_{\beta^{\prime}}a_{j\beta^{\prime},i\alpha}^{(1)}+\left(  \frac{1}{i}%
\Gamma_{1}+\gamma_{i,\alpha}^{(1)}\right)  \sigma_{i}\phi_{\alpha}%
\end{align*}
where we used (\ref{a} ). Taking $Tr(\sigma_{g}^{T}\cdot)$ of this last
equation yields%
\begin{equation}
0=\alpha_{g}\left(  \gamma_{i,\alpha}^{(0)}-i2M_{g}\Omega_{o}+D\triangledown
^{2}\right)  \sum_{\beta^{\prime}}\phi_{g\beta^{\prime}}a_{g\beta^{\prime
},i\alpha}^{(1)}+\frac{1}{i}\left[  \Gamma_{1}\right]  _{g,i}\phi_{\alpha
}+\gamma_{i,\alpha}^{(1)}\phi_{\alpha}\delta_{gi}\alpha_{g} \label{b}%
\end{equation}

where%
\begin{equation}
\left[  \Gamma_{1}\right]  _{g,i}=Tr(\sigma_{g}^{T}\Gamma_{1}\sigma
_{i})=Tr\sigma_{g}^{T}\left[  \overrightarrow{\Omega}_{1}\cdot\overrightarrow
{\sigma},\sigma_{i}\right]  \label{0004}%
\end{equation}
Making use of the orthogonality of the $\phi_{\alpha},$ and taking them to be
normalized%
\begin{equation}
\int_{V}d^{3}x\phi_{\beta}^{\ast}\phi_{\alpha}=\delta_{\beta\alpha}%
\end{equation}
we multiply (\ref{b}) by $\phi_{\beta}^{\ast}$ and integrate over the volume:%
\begin{align}
0  &  =\alpha_{g}\left(  \gamma_{i,\alpha}^{(0)}-i2M_{g}\Omega_{o}-Dk_{\beta
}^{2}\right)  a_{g\beta,i\alpha}^{(1)}+\frac{1}{i}\left\langle \beta
\right\vert \left[  \Gamma_{1}\right]  _{g,i}\left\vert \alpha\right\rangle
+\gamma_{i,\alpha}^{(1)}\delta_{gi}\delta_{\alpha\beta}\alpha_{g}\nonumber\\
0  &  =\alpha_{g}\left(  \gamma_{i,\alpha}^{(0)}-\gamma_{g,\beta}%
^{(0)}\right)  a_{g\beta,i\alpha}^{(1)}+\frac{1}{i}\left\langle \beta
\right\vert \left[  \Gamma_{1}\right]  _{g,i}\left\vert \alpha\right\rangle
+\gamma_{i,\alpha}^{(1)}\delta_{gi}\delta_{\alpha\beta}\alpha_{g} \label{c}%
\end{align}
using (\ref{13}), where%
\begin{equation}
\left\langle \beta\right\vert \left[  \Gamma_{1}\right]  _{i,j}\left\vert
\alpha\right\rangle \triangleq\int_{V}d^{3}x\phi_{\beta}^{\ast}\left[
\Gamma_{1}\right]  _{i,j}\phi_{\alpha}%
\end{equation}
We note that $k_{\alpha=0}=0,$ corresponding to a uniform distribution in the
cell, is a valid solution and we will seek the decay parameters for this mode.
Thus we put $\alpha=\beta=0,$ and $i=g$ in (\ref{c}) obtaining
\begin{align}
0  &  =\frac{1}{i}\left\langle 0\right\vert \left[  \Gamma_{1}\right]
_{i,i}\left\vert 0\right\rangle +\gamma_{i,0}^{(1)}\alpha_{i}\label{d}\\
0  &  =\gamma_{i,0}^{(1)}\\
a_{g\beta,i0}^{(1)}  &  =\frac{i\left\langle \beta\right\vert \left[
\Gamma_{1}\right]  _{g,i}\left\vert 0\right\rangle }{\alpha_{g}\left(
\gamma_{i,0}^{(0)}-\gamma_{g,\beta}^{(0)}\right)  }%
\end{align}

\bigskip The matrix element in (\ref{d}) is seen to be zero for perturbing
fields with a volume average of zero.

Now we use (\ref{18}) to evaluate the second order corrections%
\begin{align}
0  &  =\left(  \gamma_{i,\alpha}^{(0)}+\frac{1}{i}\Gamma_{o}+D\triangledown
^{2}\right)  \sum_{j\beta^{\prime}}\sigma_{j}\phi_{\beta^{\prime}}%
a_{j\beta^{\prime},i\alpha}^{(2)}+\nonumber\\
&  \left(  \frac{1}{i}\Gamma_{1}+\gamma_{i,\alpha}^{(1)}\right)  \sum
_{j\beta^{\prime}}\sigma_{j}\phi_{\beta^{\prime}}a_{j\beta^{\prime},i\alpha
}^{(1)}+\gamma_{i,\alpha}^{(2)}\sigma_{i}\phi_{\alpha}%
\end{align}
Again taking $Tr\left(  \sigma_{g}^{\ast}\cdot\right)  $ of this equation
\begin{align}
0  &  =\alpha_{g}\left(  \gamma_{i,\alpha}^{(0)}+\frac{2}{i}M_{g}\Omega
_{o}+D\triangledown^{2}\right)  \sum_{\beta^{\prime}}\phi_{\beta^{\prime}%
}a_{g\beta^{\prime},i\alpha}^{(2)}+\sum_{j\beta^{\prime}}\frac{1}{i}\left[
\Gamma_{1}\right]  _{g,j}\phi_{\beta^{\prime}}a_{j\beta^{\prime},i\alpha
}^{(1)}+..\nonumber\\
&  +\left(  \gamma_{i,\alpha}^{(1)}\sum_{\beta^{\prime}}\phi_{\beta^{\prime}%
}a_{g\beta^{\prime},i\alpha}^{(1)}+\gamma_{i,\alpha}^{(2)}\delta_{gi}%
\phi_{\alpha}\right)  \alpha_{g}\\
0  &  =\alpha_{g}\left(  \gamma_{i,\alpha}^{(0)}-\gamma_{g,\beta}%
^{(0)}\right)  a_{g\beta,i\alpha}^{(2)}+\sum_{j,\beta^{\prime}}\frac{1}%
{i}\left\langle \beta\right\vert \left[  \Gamma_{1}\right]  _{g,j}\left\vert
\beta^{\prime}\right\rangle a_{j\beta^{\prime},i\alpha}^{(1)}+\alpha_{g}%
\gamma_{i,\alpha}^{(1)}a_{g\beta,i\alpha}^{(1)}+..\nonumber\\
&  +\gamma_{i,\alpha}^{(2)}\delta_{gi}\delta_{\alpha\beta}\alpha_{g}%
\end{align}
where the last result comes from multiplying by $\phi_{\beta}^{\ast}$ and
integrating over volume. Now taking $\alpha=\beta=0,i=g,$ we find%
\begin{align}
0  &  =\sum_{j,\beta^{\prime}}\frac{1}{i}\left\langle 0\right\vert \left[
\Gamma_{1}\right]  _{i,j}\left\vert \beta^{\prime}\right\rangle a_{j\beta
^{\prime},i0}^{(1)}+\gamma_{i,0}^{(2)}\alpha_{i}\nonumber\\
\alpha_{i}\gamma_{i,0}^{(2)}  &  =-\sum_{j,\beta^{\prime}}\left\langle
0\right\vert \left[  \Gamma_{1}\right]  _{i,j}\left\vert \beta^{\prime
}\right\rangle \frac{\left\langle \beta^{\prime}\right\vert \left[  \Gamma
_{1}\right]  _{j,i}\left\vert 0\right\rangle }{\left(  \gamma_{i,0}%
^{(0)}-\gamma_{j,\beta^{\prime}}^{(0)}\right)  \alpha_{j}} \label{aa}%
\end{align}
Our derivation has followed the method of time independent
Rayleigh-Schroedinger perturbation theory. The 'states' are characterized by
two 'quantum numbers' a spin index $i$ and a spatial index $\alpha,$ which can
stand for 3 indices, which appear when we solve the diffusion equation in 3 dimensions.

\subsection{Calculation of relaxation times, relation to McGregor's result}

We begin by evaluating (\ref{aa}) for $i=0.$ Since $\sigma_{o}=\sigma_{z}$
this will be equal to $1/T_{1}$. We have then $\left(  \phi_{\alpha=0}%
=1/\sqrt{V}\right)  $%
\begin{equation}
\left\langle \beta^{\prime}\right\vert \left[  \Gamma_{1}\right]
_{j,i}\left\vert 0\right\rangle =\frac{1}{\sqrt{V}}\int d^{3}r\phi
_{\beta^{\prime}}\left(  \overrightarrow{r}\right)  \left[  \Gamma_{1}\right]
_{j,0}%
\end{equation}
We write
\begin{align}
\overrightarrow{\Omega}_{1}\cdot\overrightarrow{\sigma}  &  =\left(  \left[
\Omega_{1}\right]  _{+}\sigma_{-}+\left[  \Omega_{1}\right]  _{-}\sigma
_{+}\right)  +\left[  \Omega_{1}\right]  _{z}\sigma_{z}\\
\left[  \overrightarrow{\Omega}_{1}\cdot\overrightarrow{\sigma},\sigma
_{o}\right]   &  =2\left(  \left[  \Omega_{1}\right]  _{+}\sigma_{-}-\left[
\Omega_{1}\right]  _{-}\sigma_{+}\right) \\
\left[  \Gamma_{1}\right]  _{j,0}  &  =Tr\sigma_{j}^{T}\left[  \overrightarrow
{\Omega}_{1}\cdot\overrightarrow{\sigma},\sigma_{o}\right] \\
\left[  \Gamma_{1}\right]  _{-,0}  &  =2\left[  \Omega_{1}\right]
_{+}\label{001a}\\
\left[  \Gamma_{1}\right]  _{+,0}  &  =-2\left[  \Omega_{1}\right]  _{-}
\label{001b}%
\end{align}
Thus%
\begin{align}
\gamma_{i=0,0}^{(2)}  &  =\frac{1}{T_{1}}=-\frac{1}{\alpha_{o}}\sum
_{j,\beta^{\prime}}\left\langle 0\right\vert \left[  \Gamma_{1}\right]
_{0,j}\left\vert \beta^{\prime}\right\rangle \frac{\left\langle \beta^{\prime
}\right\vert \left[  \Gamma_{1}\right]  _{j,0}\left\vert 0\right\rangle
}{\left(  \gamma_{0,0}^{(0)}-\gamma_{j,\beta^{\prime}}^{(0)}\right)
\alpha_{j}}\nonumber\\
&  =\frac{4}{V\alpha_{o}}\int\int\left[  \Omega_{1}\left(  \overrightarrow
{r^{\prime}}\right)  \right]  _{-}\left[  \Omega_{1}\left(  \overrightarrow
{r}\right)  \right]  _{+}\sum_{\beta^{\prime}}\ \left(  \frac{\phi
_{\beta^{\prime}}\left(  \overrightarrow{r^{\prime}}\right)  \phi
_{\beta^{\prime}}\left(  \overrightarrow{r}\right)  }{(Dk_{\beta^{\prime}}%
^{2}-i2\Omega_{o})}\right)  d^{3}r^{\prime}d^{3}r\nonumber\\
&  +cc\nonumber\\
\frac{1}{T_{1}}  &  =\frac{4}{V}\operatorname{Re}\int\int\left[  \Omega
_{1}\left(  \overrightarrow{r^{\prime}}\right)  \right]  _{-}\left[
\Omega_{1}\left(  \overrightarrow{r}\right)  \right]  _{+}\sum_{\beta^{\prime
}}\ \left(  \frac{\phi_{\beta^{\prime}}\left(  \overrightarrow{r^{\prime}%
}\right)  \phi_{\beta^{\prime}}\left(  \overrightarrow{r}\right)  }%
{(Dk_{\beta^{\prime}}^{2}-i2\Omega_{o})}\right)  d^{3}r^{\prime}%
d^{3}r\nonumber\\
&  \label{111a}%
\end{align}

Now the Green's function for the diffusion equation can be written (see Morse
and Feshbach, [\cite{MandF}] chapter 7)%
\begin{equation}
G\left(  \overrightarrow{r},t|\overrightarrow{r}^{\prime},t^{\prime}\right)
=u\left(  t-t^{\prime}\right)  \sum_{\beta}\phi_{\beta}\left(  \overrightarrow
{r^{\prime}}\right)  \phi_{\beta}\left(  \overrightarrow{r}\right)
e^{-Dk_{\beta}^{2}(t-t^{\prime})}%
\end{equation}
with $u\left(  t\right)  ,$ the unit step function. Then the time Fourier
transform is $\left(  \tau=t-t^{\prime}\right)  $%
\begin{align}
\widetilde{G}\left(  \overrightarrow{r},\overrightarrow{r}^{\prime}%
,\omega\right)   &  =\int_{0}^{\infty}d\tau e^{i\omega\tau}G\left(
\overrightarrow{r},t|\overrightarrow{r}^{\prime},t^{\prime}\right) \\
&  =\int_{0}^{\infty}d\tau e^{i\omega\tau}\sum_{\beta}\phi_{\beta}\left(
\overrightarrow{r^{\prime}}\right)  \phi_{\beta}\left(  \overrightarrow
{r}\right)  e^{-Dk_{\beta}^{2}\tau}\\
&  =\sum_{\beta^{\prime}}\ \left(  \frac{\phi_{\beta^{\prime}}\left(
\overrightarrow{r^{\prime}}\right)  \phi_{\beta^{\prime}}\left(
\overrightarrow{r}\right)  }{(Dk_{\beta^{\prime}}^{2}-i\omega)}\right)
\label{1a}%
\end{align}
Comparing to the sum in (\ref{111a}) we see that we can write $\left(
\alpha_{0}=2\right)  $%
\begin{align}
\frac{1}{T_{1}}  &  =\operatorname{Re}\gamma_{i=0,0}^{(2)}\nonumber\\
&  =\frac{4\operatorname{Re}}{V}\int\int\left[  \Omega_{1}\left(
\overrightarrow{r^{\prime}}\right)  \right]  _{-}\left[  \Omega_{1}\left(
\overrightarrow{r}\right)  \right]  _{+}\widetilde{G}\left(  \overrightarrow
{r},\overrightarrow{r}^{\prime},2\Omega_{o}\right)  d^{3}r^{\prime}%
d^{3}r\nonumber\\
&  =4\frac{\operatorname{Re}}{V}\int_{0}^{\infty}d\tau e^{i2\Omega_{o}\tau
}\int\int\left[  \Omega_{1}\left(  \overrightarrow{r^{\prime}}\right)
\right]  _{-}\left[  \Omega_{1}\left(  \overrightarrow{r}\right)  \right]
_{+}G\left(  \overrightarrow{r},t|\overrightarrow{r}^{\prime},t^{\prime
}\right)  d^{3}r^{\prime}d^{3}r\nonumber\\
&  =4\operatorname{Re}\int_{0}^{\infty}d\tau e^{i\omega_{o}\tau}\int
\int\left[  \Omega_{1}\left(  \overrightarrow{r^{\prime}}\right)  \right]
_{-}\left[  \Omega_{1}\left(  \overrightarrow{r}\right)  \right]  _{+}G\left(
\overrightarrow{r},t|\overrightarrow{r}^{\prime},t^{\prime}\right)
p_{o}(r^{\prime},t^{\prime})d^{3}r^{\prime}d^{3}r\nonumber\\
&  \label{111b}%
\end{align}
Where $\omega_{o}=\gamma B_{o}$ and $p_{o}(r^{\prime},t^{\prime})=1/V$ is the
uniform density of magnetization. Then the joint probability distribution of
an atom being at $\overrightarrow{r}$ at time $t$ and being at
$\overrightarrow{r}^{\prime}$ at time $t^{\prime}$ is $G\left(
\overrightarrow{r},t|\overrightarrow{r}^{\prime},t^{\prime}\right)
p_{o}(r^{\prime},t^{\prime})$ and we see that (following McGregor's notation)%
\begin{align}
\frac{1}{T_{1}}  &  =\gamma_{i=0,0}^{(2)}=4\operatorname{Re}\int_{0}^{\infty
}d\tau e^{i\omega_{o}\tau}\left\langle \left[  \Omega_{1}\left(  t\right)
\right]  _{-}\left[  \Omega_{1}\left(  t+\tau\right)  \right]  _{+}%
\right\rangle \nonumber\\
&  =4\int_{0}^{\infty}d\tau e^{i\omega_{o}\tau}\left\langle \left[  \Omega
_{1}\left(  t\right)  \right]  _{x}\left[  \Omega_{1}\left(  t+\tau\right)
\right]  _{x}+\left[  \Omega_{1}\left(  t\right)  \right]  _{y}\left[
\Omega_{1}\left(  t+\tau\right)  \right]  _{y}\right\rangle \nonumber\\
&  =\frac{\gamma^{2}}{2}\int_{-\infty}^{\infty}d\tau e^{i\omega_{o}\tau
}\left\langle \left[  B_{1}\right]  _{x}\left[  B_{1}\left(  t+\tau\right)
\right]  _{x}+\left[  B_{1}\left(  t\right)  \right]  _{y}\left[  B_{1}\left(
t+\tau\right)  \right]  _{y}\right\rangle \label{T1}%
\end{align}
This is the result of the Redfield theory given as equation (9) in McGregor.

To calculate $T_{2}$ we have to evaluate $\gamma_{i=+,0}^{(2)}.$%
\begin{equation}
\gamma_{i=+,0}^{(2)}=-\sum_{j,\beta^{\prime}}\left\langle 0\right\vert \left[
\Gamma_{1}\right]  _{+,j}\left\vert \beta^{\prime}\right\rangle \frac
{\left\langle \beta^{\prime}\right\vert \left[  \Gamma_{1}\right]
_{j,+}\left\vert 0\right\rangle }{\left(  \gamma_{+,0}^{(0)}-\gamma
_{j,\beta^{\prime}}^{(0)}\right)  \alpha_{j}}%
\end{equation}
The non-zero matrix elements are
\begin{align}
\left[  \Gamma_{1}\right]  _{+,+}  &  =2\left[  \Omega_{1}\right]
_{z}\label{001c}\\
\left[  \Gamma_{1}\right]  _{o,+}  &  =-2\left[  \Omega_{1}\right]  _{+}
\label{001d}%
\end{align}
so that%
\begin{align}
\gamma_{i=+,0}^{(2)}  &  =\frac{1}{T_{2}}=-\operatorname{Re}\sum
_{\beta^{\prime}}\left[
\begin{array}
[c]{c}%
\left\langle 0\right\vert \left[  \Gamma_{1}\right]  _{+,+}\left\vert
\beta^{\prime}\right\rangle \frac{\left\langle \beta^{\prime}\right\vert
\left[  \Gamma_{1}\right]  _{+,+}\left\vert 0\right\rangle }{\left(
\gamma_{+,0}^{(0)}-\gamma_{+,\beta^{\prime}}^{(0)}\right)  }+\\
\left\langle 0\right\vert \left[  \Gamma_{1}\right]  _{+,0}\left\vert
\beta^{\prime}\right\rangle \frac{\left\langle \beta^{\prime}\right\vert
\left[  \Gamma_{1}\right]  _{0,+}\left\vert 0\right\rangle }{2\left(
\gamma_{+,0}^{(0)}-\gamma_{0,\beta^{\prime}}^{(0)}\right)  }%
\end{array}
\right] \\
&  =-\operatorname{Re}\sum_{\beta^{\prime}}\left[
\begin{array}
[c]{c}%
4\frac{\left\langle 0\right\vert \left[  \Omega_{1}\right]  _{z}\left\vert
\beta^{\prime}\right\rangle \left\langle \beta^{\prime}\right\vert \left[
\Omega_{1}\right]  _{z}\left\vert 0\right\rangle }{\left(  \gamma_{+,0}%
^{(0)}-\gamma_{+,\beta^{\prime}}^{(0)}\right)  }+\\
4\frac{\left\langle 0\right\vert \left[  \Omega_{1}\right]  _{-}\left\vert
\beta^{\prime}\right\rangle \left\langle \beta^{\prime}\right\vert \left[
\Omega_{1}\right]  _{+}\left\vert 0\right\rangle }{2\left(  \gamma_{+,0}%
^{(0)}-\gamma_{0,\beta^{\prime}}^{(0)}\right)  }%
\end{array}
\right] \nonumber\\
&  =4\operatorname{Re}\sum_{\beta^{\prime}}\left[  \frac{\left\langle
0\right\vert \left[  \Omega_{1}\right]  _{z}\left\vert \beta^{\prime
}\right\rangle \left\langle \beta^{\prime}\right\vert \left[  \Omega
_{1}\right]  _{z}\left\vert 0\right\rangle }{\left(  Dk_{\beta^{\prime}}%
^{2}\right)  }+\frac{\left\langle 0\right\vert \left[  \Omega_{1}\right]
_{-}\left\vert \beta^{\prime}\right\rangle \left\langle \beta^{\prime
}\right\vert \left[  \Omega_{1}\right]  _{+}\left\vert 0\right\rangle
}{2\left(  Dk_{\beta^{\prime}}^{2}-i\omega_{o}\right)  }\right] \nonumber\\
\frac{1}{T_{2}}  &  =\frac{4\operatorname{Re}}{V}\int\int\left[  \Omega
_{1}\left(  \overrightarrow{r^{\prime}}\right)  \right]  _{z}\left[
\Omega_{1}\left(  \overrightarrow{r}\right)  \right]  _{z}\sum_{\beta^{\prime
}}\ \left(  \frac{\phi_{\beta^{\prime}}\left(  \overrightarrow{r^{\prime}%
}\right)  \phi_{\beta^{\prime}}\left(  \overrightarrow{r}\right)  }%
{Dk_{\beta^{\prime}}^{2}}\right)  d^{3}r^{\prime}d^{3}r+\nonumber\\
&  \frac{2\operatorname{Re}}{V}\int\int\left[  \Omega_{1}\left(
\overrightarrow{r^{\prime}}\right)  \right]  _{-}\left[  \Omega_{1}\left(
\overrightarrow{r}\right)  \right]  _{+}\sum_{\beta^{\prime}}\ \left(
\frac{\phi_{\beta^{\prime}}\left(  \overrightarrow{r^{\prime}}\right)
\phi_{\beta^{\prime}}\left(  \overrightarrow{r}\right)  }{(Dk_{\beta^{\prime}%
}^{2}-i\omega_{o})}\right)  d^{3}r^{\prime}d^{3}r\nonumber\\
&  \label{11}%
\end{align}

\bigskip From (\ref{1a}) we write%
\begin{align}
\widetilde{G}\left(  \overrightarrow{r},\overrightarrow{r}^{\prime}%
,\omega=0\right)   &  =\sum_{\beta^{\prime}}\ \left(  \frac{\phi
_{\beta^{\prime}}\left(  \overrightarrow{r^{\prime}}\right)  \phi
_{\beta^{\prime}}\left(  \overrightarrow{r}\right)  }{Dk_{\beta^{\prime}}^{2}%
}\right) \nonumber\\
&  =\int_{0}^{\infty}d\tau G\left(  \overrightarrow{r},t|\overrightarrow
{r}^{\prime},t^{\prime}\right)  \label{111c}%
\end{align}
so that the first term in (\ref{11}) can be written%
\begin{align}
&  \frac{\gamma^{2}\operatorname{Re}}{2}\int_{-\infty}^{\infty}d\tau\int
\int\left[  B_{1}\left(  \overrightarrow{r^{\prime}}\right)  \right]
_{z}\left[  B_{1}\left(  \overrightarrow{r}\right)  \right]  _{z}G\left(
\overrightarrow{r},t|\overrightarrow{r}^{\prime},t+\tau\right)  p_{o}\left(
r^{\prime}\right)  d^{3}r^{\prime}d^{3}r\nonumber\\
&  =\frac{\gamma^{2}}{2}\int_{-\infty}^{\infty}d\tau\left\langle \left[
B_{1}\left(  t\right)  \right]  _{z}\left[  B_{1}\left(  t+\tau\right)
\right]  _{z}\right\rangle
\end{align}
in agreement with the second term in equ (10) of McGregor.

From (\ref{111a}) we see that the second term is $1/2T_{1}$ so that equation
(\ref{11}) is equivalent to%
\begin{equation}
\frac{1}{T_{2}}=\frac{1}{2T_{1}}+\frac{\gamma^{2}}{2}\int_{-\infty}^{\infty
}d\tau\left\langle \left[  B_{1}\left(  t\right)  \right]  _{z}\left[
B_{1}\left(  t+\tau\right)  \right]  _{z}\right\rangle \label{XX}%
\end{equation}
which is equivalent to equation (10) of \cite{McG}.

\section{Appendix B, spin relations and matrix elements}

\bigskip$\left[  \sigma_{1},\sigma_{z}\right]  =-2\sigma_{1}\qquad\left[
\sigma_{2},\sigma_{z}\right]  =2\sigma_{2}\qquad\left[  \sigma_{1},\sigma
_{2}\right]  =\sigma_{z}$%

\begin{align}
\overrightarrow{\Omega}_{1}\cdot\overrightarrow{\sigma}  &  =\left[
\Omega_{1}\right]  _{+}\sigma_{-}+\left[  \Omega_{1}\right]  _{-}\sigma
_{+}+\left[  \Omega_{1}\right]  _{z}\sigma_{z}\qquad\ \left(  \sigma_{\pm
}=\frac{1}{2}\left(  \sigma_{x}\pm i\sigma_{y}\right)  \right) \\
\left[  \overrightarrow{\Omega}_{1}\cdot\overrightarrow{\sigma},\sigma
_{z}\right]   &  =2\left(  \left[  \Omega_{1}\right]  _{+}\sigma_{-}-\left[
\Omega_{1}\right]  _{-}\sigma_{+}\right) \\
\left[  \Gamma_{1}\right]  _{g.i}  &  =Tr\left(  \sigma_{g}^{T}\left[
\overrightarrow{\Omega}_{1}\cdot\overrightarrow{\sigma},\sigma_{i}\right]
\right) \\
\left[  \Gamma_{1}\right]  _{+.z}  &  =-2\left[  \Omega_{1}\right]  _{-}%
\qquad\left[  \Gamma_{1}\right]  _{-.z}=2\left[  \Omega_{1}\right]  _{+}\\
\left[  \overrightarrow{\Omega}_{1}\cdot\overrightarrow{\sigma},\sigma
_{+}\right]   &  =-\left[  \Omega_{1}\right]  _{+}\sigma_{z}+\left[
\Omega_{1}\right]  _{z}2\sigma_{+}\\
\left[  \Gamma_{1}\right]  _{z,+}  &  =-2\left[  \Omega_{1}\right]  _{+}%
\qquad\left[  \Gamma_{1}\right]  _{+,+}=2\left[  \Omega_{1}\right]  _{z}\\
\left[  \overrightarrow{\Omega}_{1}\cdot\overrightarrow{\sigma},\sigma
_{-}\right]   &  =\left[  \Omega_{1}\right]  _{-}\sigma_{z}-\left[  \Omega
_{1}\right]  _{z}2\sigma_{-}\\
\left[  \Gamma_{1}\right]  _{z,-}  &  =2\left[  \Omega_{1}\right]  _{-}%
\qquad\left[  \Gamma_{1}\right]  _{-,-}=-2\left[  \Omega_{1}\right]  _{z}%
\end{align}
Note
\begin{align}
\sum_{n=0}^{\infty}\frac{1}{\left(  2n+1\right)  ^{4}}  &  =\frac{1}{96}%
\pi^{4}\label{pow4}\\
\sum_{n=0}^{\infty}\frac{1}{\left(  2n+1\right)  ^{6}}  &  =\allowbreak
\frac{1}{960}\pi^{6}\label{pow6}\\
\sum_{n=0}^{\infty}\frac{1}{\left(  2n+1\right)  ^{2}}  &  =\allowbreak
\frac{1}{8}\pi^{2} \label{pow2}%
\end{align}

\bigskip%

\begin{figure}
[ptb]
\begin{center}
\includegraphics[
trim=0.000000in 0.000000in 0.000000in -0.008000in,
natheight=7.999600in,
natwidth=14.222600in,
height=4.3624in,
width=5.2239in
]%
{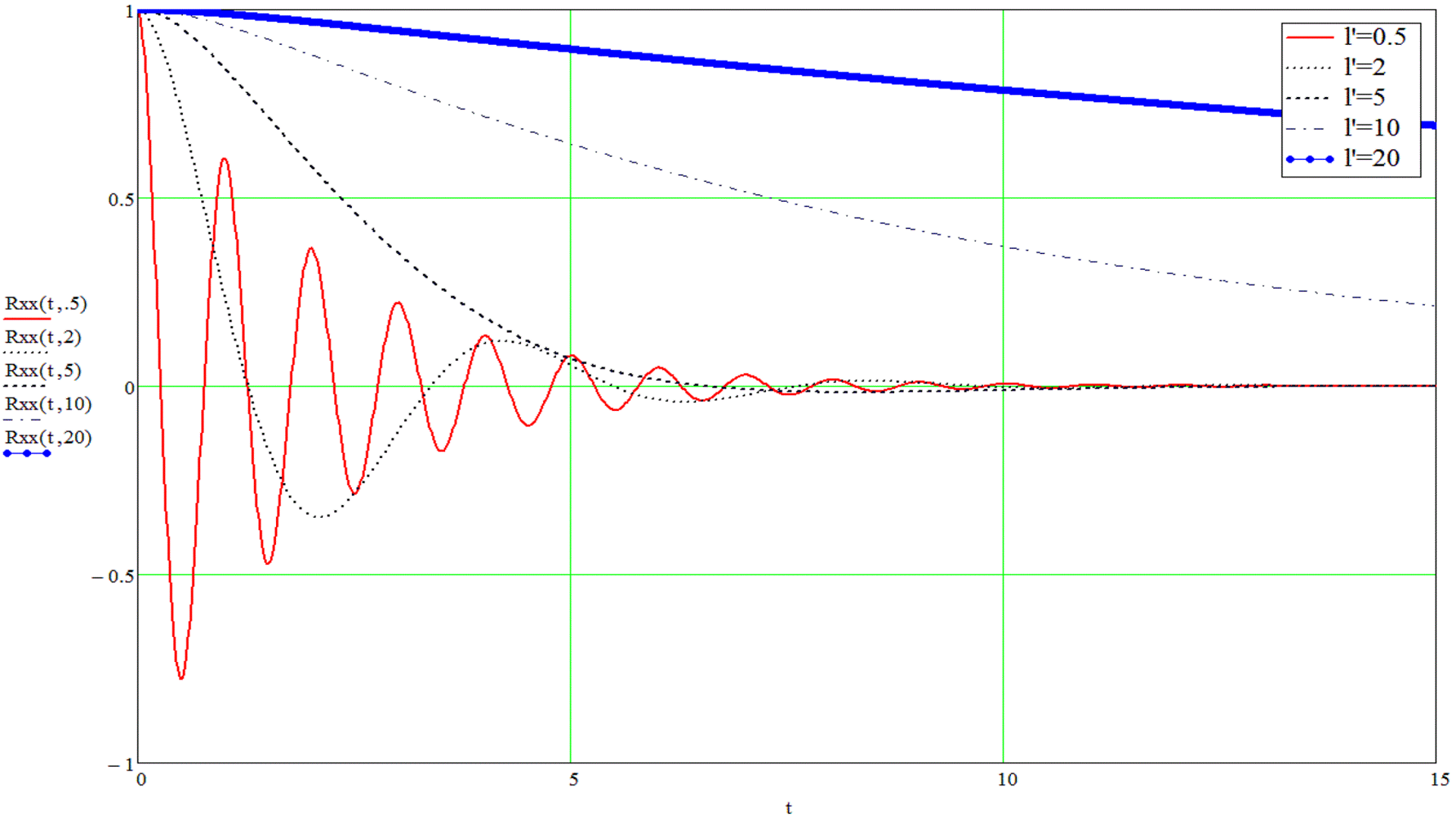}%
\caption{The autocorrelation function for particle position \ for particles
moving in a rectangular box as a function of dimensionless delay time,
$\tau^{\prime},$ with $l^{\prime}=L_{x}/\lambda$, where $L_{x}$ is the length
of the cell in the $x$ direction and $\lambda$ is the mean free path between
collisions, as a parameter.}%
\label{Figure 1}%
\end{center}
\end{figure}
%

\begin{figure}
[ptb]
\begin{center}
\includegraphics[
natheight=7.999600in,
natwidth=14.222600in,
height=4.5717in,
width=5.2239in
]%
{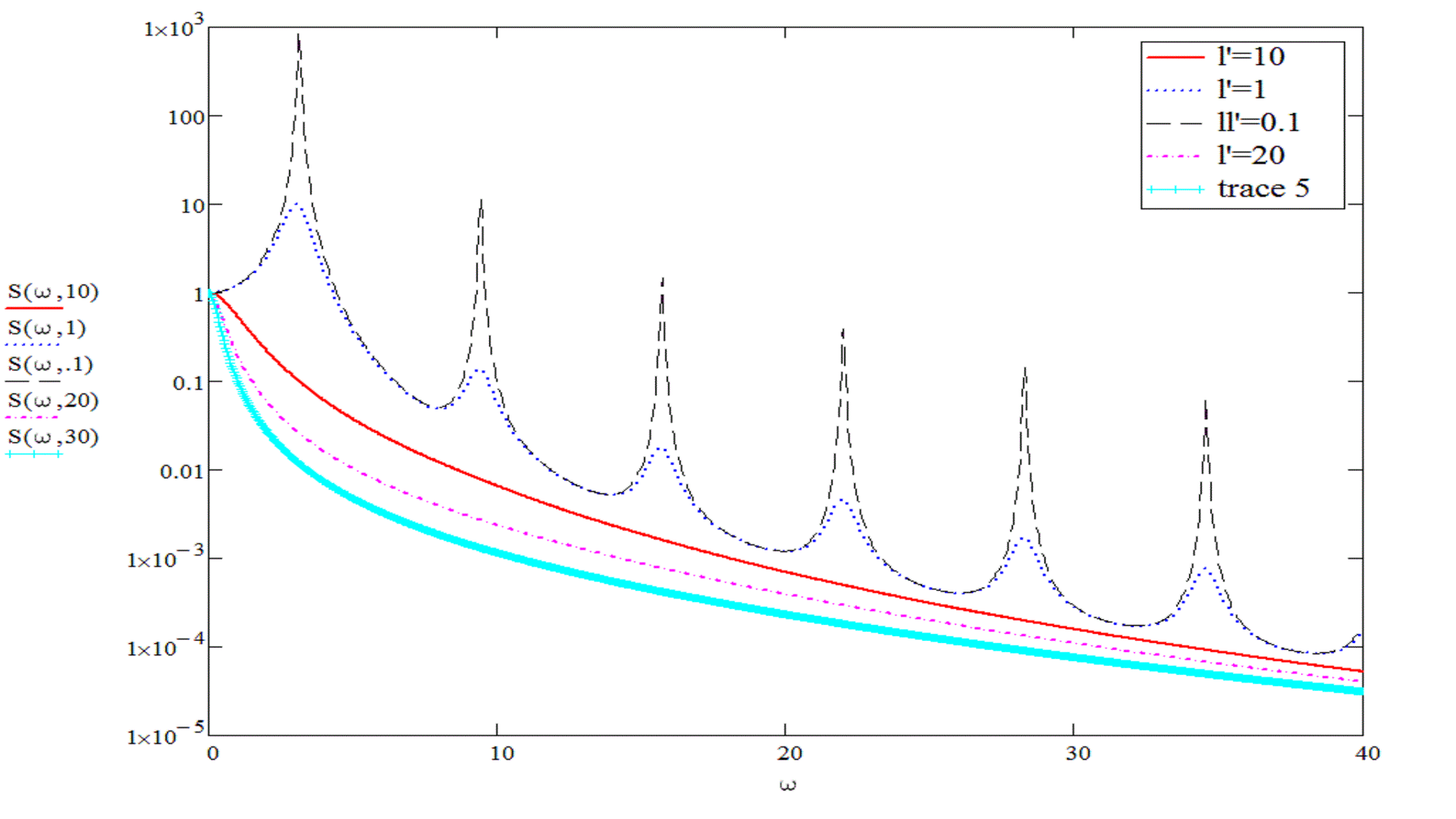}%
\caption{Frequency spectrum of the auto-correlation function of fig.1 as a
function of reduced frequency and $l^{\prime}$.}%
\label{Figure 2}%
\end{center}
\end{figure}

\end{document}